\documentclass[useAMS,usenatbib]{mn2e}
\usepackage{graphicx}
\title[Lithium Abundances in Stars with Planets]{Parent Stars of Extrasolar Planets. IX. Lithium Abundances}
\author[G. Gonzalez]{Guillermo Gonzalez\thanks{E-mail:
gonzog@iastate.edu} \\
Iowa State University, Department of Physics and Astronomy, Ames, 
IA 50011}
\begin{document}

\date{Accepted ??. Received ??; in original form ??}

\pagerange{\pageref{firstpage}--\pageref{lastpage}} \pubyear{??}

\maketitle

\label{firstpage}

\begin{abstract}
We compare the Li abundances of a sample of stars with planets discovered with the Doppler method to a sample of stars without detected planets. We prepared the samples by combining the Li abundances reported in several recent studies in a consistent way. Our results confirm recent claims that the Li abundances of stars with planets are smaller than those of stars without planets near the solar temperature. We also find that the vsini and $R^{'}_{\rm HK}$ anomalies correlate with the Li abundance anomalies. These results suggest that planet formation processes have altered the rotation and Li abundances of stars that host Doppler detected planets. We encourage others to test these findings with additional observations of Li in stars with temperatures between 5600 and 6200 K.
\end{abstract}

\section{Introduction}

Stellar Li abundances are at once very informative and difficult to interpret. This follows from the relative delicacy of Li nuclei in the shallow surface layers of stars, where they are destroyed via $(p,\alpha)$ reactions when they are mixed to regions with warm protons. Li abundances in dwarf star photospheres are observed to correlate with effective temperature (T$_{\rm eff}$), age, rotation, binarity and metallicity. However, even within a single coeval population (such as the open cluster M67) variations in Li abundance are observed among stars that otherwise appear identical \citep{rand06}.

Some have suggested that the presence of a protoplanetary disk is the missing parameter that accounts for the observed spread in Li abundances among similar stars. The the surface Li abundance of a star could be altered via accretion of protoplanetary disk material \citep{gg98} or via a change in its rotation \citep{chen06,tak07}. Thus, while Li has the potential to serve as a useful probe of stellar and planetary processes, their effects on Li abundances are difficult to disentangle given our present level of understanding.

Nevertheless, several studies have attempted to isolate the effects of planets on Li abundance. \citet{gl00} first suggested that stars with planets (SWPs), when corrected for simple linear trends with T$_{\rm eff}$, metallicity\footnote{In the following we use either [Fe/H] or [M/H] as a proxy for stellar metallicity.} and age, display smaller Li abundances than field stars. \citet{ryan00} examined the Li abundance trends more carefully and concluded that any possible differences were not significant. \citet{gg01}, employing a larger sample, agreed with his conclusion.

\citet{is04} revisited this topic and reported a significant depletion of Li among SWPs relative to a comparison star sample, but only in the T$_{\rm eff}$ range 5600 to 5850 K. \citet{tak05} largely confirmed their findings for the T$_{\rm eff}$ range 5800 to 5900 K. \citet{chen06}, restricting their attention to T$_{\rm eff} =$ 5600 to 5900 K, also confirmed the conclusions of \citet{is04}. However, \citet{luck06}, employing a larger comparison star sample, did not find a significant difference between SWPs and a comparison sample. They attribute the Li abundance difference found by \citet{is04} to a systematic difference in the temperature scales in their study and the study of \citet{chen01}, the results of which they had used to supplement their small comparison sample. Most recently, \citet{tak07} measured Li in 118 nearby solar analogs. While they included only a few SWPs in their study, they again concluded that SWPs tend to have smaller Li abundances.

The purpose of the present study is to resolve the conflicting findings concerning Li abundances in SWPs. It continues our series of studies on the chemical abundances of nearby SWPs (for a summary of previous papers, see \citet{gl07}). In \citet{gl07} we combined the chemical abundance data from multiple studies in a consistent way and compared the chemical abundances of SWPs and comparison stars for several elements; we did not include Li in the comparison, because it requires a fundamentally different analysis. We employ similar methods in the present study. In Section 2 we present new samples of SWPs and comparison stars formed by combining the Li abundance results from multiple studies. In Section 3 we use these new samples to determine whether SWPs have different Li abundances than stars without detected planets. We also examine vsini and the $R^{'}_{\rm HK}$ activity index in SWPs. We discuss our findings within the context of  planet formation models in Section 4. We present our conclusions in Section 5.

\section{Combining Samples}

In \citet{gl07} we presented the results of our chemical abundance analyses of 18 elements (including Li) in 31 SWPs. We also calculated corrections needed to place our data and the data from other similar studies on the same abundance scale. This procedure allowed us to construct larger and more accurate SWP and comparison star samples, which we employed to compare C, O, Na, Mg, Al, Si, Ca, Sc, Ti and Ni abundances. We concluded that there is evidence for significant differences between the two samples for some elements.

We adopted a similar procedure for Li. We begin with the Li abundance values listed in Table 1 of \citet{gl07} for 31 SWPs. To these data we added Li abundance values (and upper limits) of other SWPs from our previous papers. We excluded SWPs with T$_{\rm eff} < 5000$ K and $\log$ g $< 4.0$. The derived stellar properties tend to be less accurate for cooler stars, and few stars with T$_{\rm eff} < 5000$ K have detectable Li. The limit on surface gravity restricts the sample to main sequence stars; a star that has evolved off the main sequence onto the subgiant or giant branch undergoes dramatic changes in its internal structure that results in changes in the surface Li abundance. Having applied these limits, the total number of SWPs with Li abundances (and upper limits) from our studies comes to 52.

Next, we compiled Li abundances (and upper limits) for SWPs with T$_{\rm eff} > 5000$ K and $\log$ g $> 4.0$ from the following studies (number of SWPs retained from each study is also indicated): \citet{is04} -- 83, \citet{luck06} -- 48, \citet{tak05} -- 27, \citet{tak07} -- 5. The same studies include Li abundances for stars without planets; the totals (after applying the same limits) are: \citet{is04} -- 34, \citet{luck06} -- 112, \citet{tak05} -- 98, \citet{tak07} -- 113.\footnote{Note, we removed HD 33636 from the list of SWPs, given the finding by \citet{bean07} that its companion has a mass in the stellar regime. The SWP totals listed for each study include HD 33636, but we moved it to the stars without planets category prior to conducting the analysis in the next section.} We selected these studies for analysis because they have similar temperature scales, and their data are of comparable quality.

One complication that we did not have to consider for the elements examined in \citet{gl07} is the presence of upper limits in the abundance data. The minimum detectable Li abundance varies among the studies we employed, because their spectra have differing S/N ratios. The Li abundance values from our series of studies typically have the lowest upper limits, while those from \citet{luck06} have the highest (but the upper limits from the other studies are not much lower). We show the stars from \citet{luck06} that have Li abundance upper limits in Figure 1. We also include in the figure a straight line that matches the upper envelope of the upper limits for T$_{\rm eff} < 6200$ K; it is described by the equation $\log$ Li $= -3.41 + 0.00079167$ T$_{\rm eff}$. We will employ this line as a cutoff for selecting Li abundance values to include in our final sample preparation below.

Next, we selected the \citet{luck06} study as the reference and adjusted the data from the other studies to be consistent with it. First, we adjusted the T$_{\rm eff}$ values in each study by applying a simple linear equation calibrated using SWPs in common between it and \citet{luck06}. We then adjusted the $\log$ g values using a linear equation including a term with the adjusted T$_{\rm eff}$ values. The adjustments to [Fe/H] and Li are simple constant offsets; typical [Fe/H] and Li abundance offsets are about 0.02 and 0.05 dex, respectively.

The cutoff line shown in Figure 1 is only valid for T$_{\rm eff} < 6200$ K, since two of the stars without planets from \citet{luck06} above this temperature have upper limits much higher than predicted from the extrapolated cutoff line. One possible reason for this discrepancy is the fact that faster rotation is more common among earlier spectral type dwarfs, resulting in broader, shallower absorption lines. Therefore, in our second cut of the data, we removed stars with T$_{\rm eff} > 6250$ K; we also excluded stars with T$_{\rm eff} < 5550$ K, given the paucity of SWPs with detected Li in this range. Our upper T$_{\rm eff}$ limit also excludes stars that might display the Li dip phenomenon (see \citet{ba95}), which would further complicate our analysis below. Finally, we excluded all the stars with $\log$ Li values falling below the cutoff line (whether upper limits or detections).

\begin{figure}
  \includegraphics[width=3.5in]{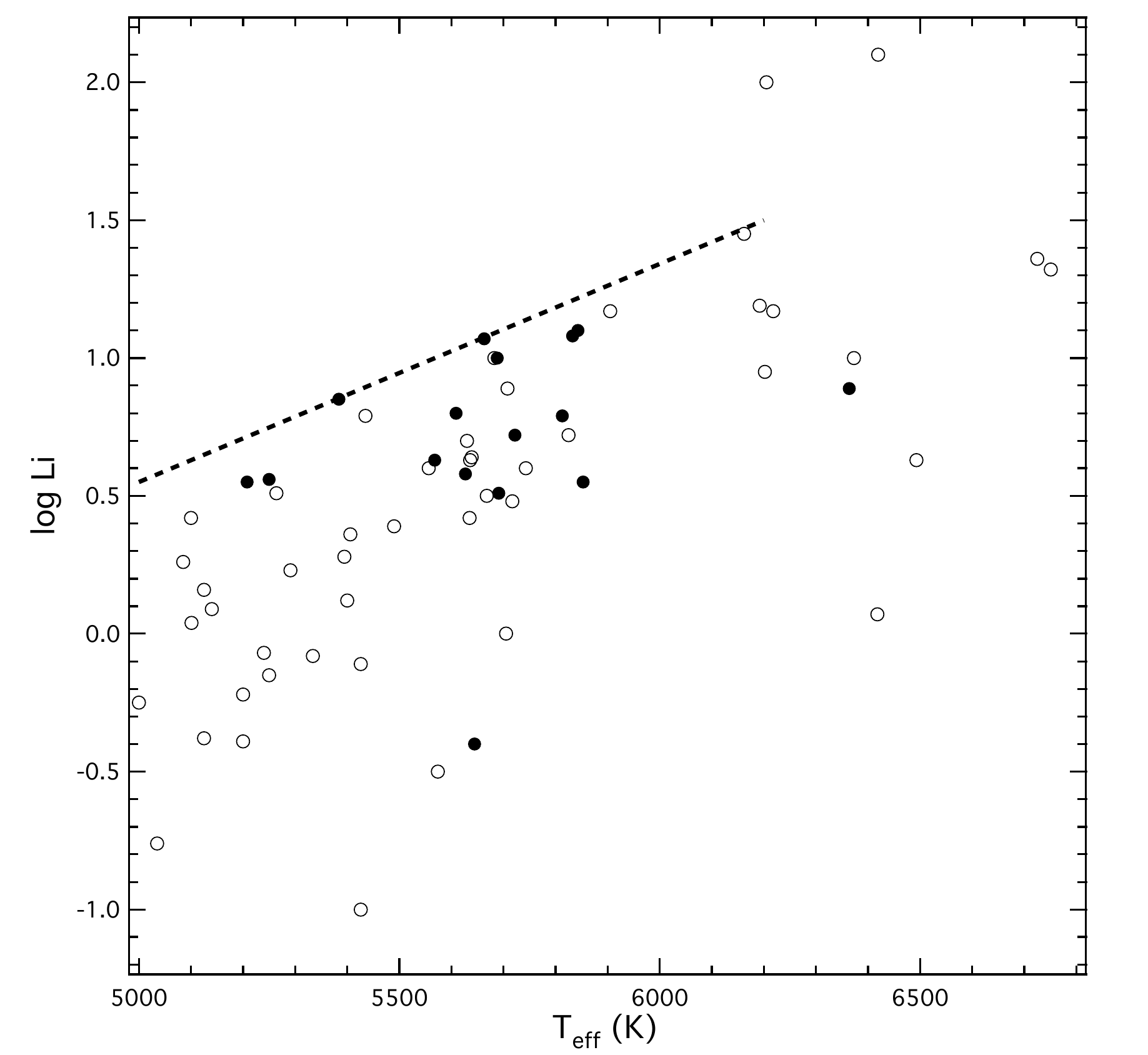}
 \caption{Li abundance upper limits versus T$_{\rm eff}$ for SWPs (dots) and stars without planets (open circles) from \citet{luck06}. The upper limit cutoff curve is shown as a dashed line.}
\end{figure}

For those stars present in multiple studies, we calculated simple averages of their parameters. We show the resulting Li abundances plotted against T$_{\rm eff}$ in Figure 2. The final samples include 37 SWPs and 147 stars lacking planets.

The corrected individual T$_{\rm eff}$ values differ from the \citet{luck06} values with a typical dispersion of about $\pm$ 70 K. The corresponding dispersion for $\log$ g is about $\pm$ 0.13 dex, while for [Fe/H] and $\log$ Li they are $\pm$ 0.03 and $\pm$ 0.08 dex, respectively. These numbers are consistent with the quoted typical measurement uncertainties in the individual studies. Of course, the standard error of the mean (s.e.m.) values for the average values of these quantities for those stars included in multiple studies are smaller.

\begin{figure}
  \includegraphics[width=3.5in]{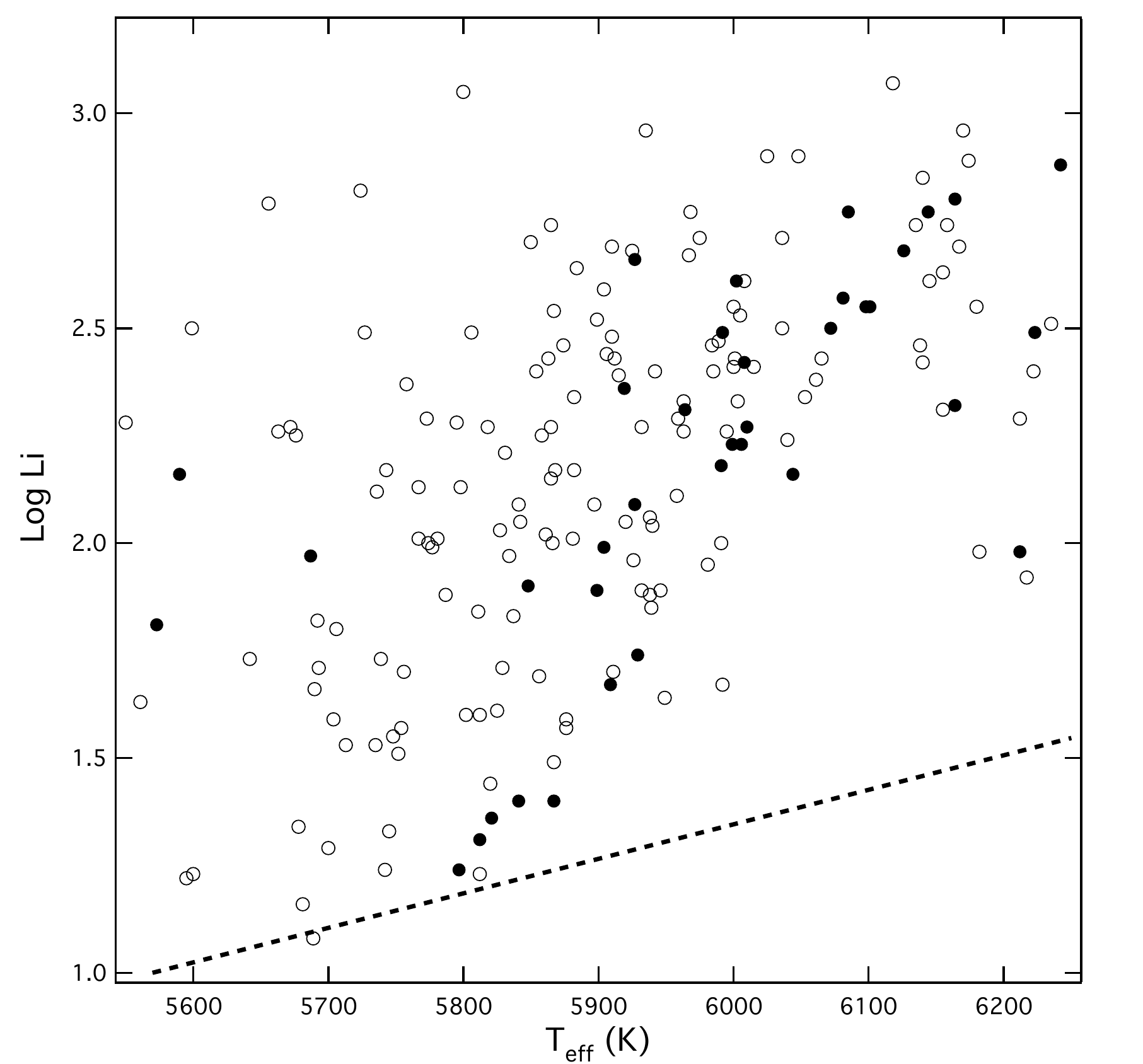}
 \caption{Final adopted Li abundances versus T$_{\rm eff}$. Symbols have the same meanings as in Figure 1.}
\end{figure}

\section{Comparing Samples}

Objective comparisons between the Li abundances of SWPs and comparison stars have proven to be difficult. Some studies have done nothing more than simple visual comparisons on a plot like Figure 2, while others have attempted to correct for differences in T$_{\rm eff}$, [Fe/H] and chromospheric activity (e.g., \citet{gl00}). We propose a new approach to comparing these samples.

Observations of stars in open clusters and in the field indicate that Li abundances are most sensitive to T$_{\rm eff}$ and age (see \citet{cut03,tak07} and references therein). In addition, the rate of Li destruction in a star's envelope should be sensitive to [Fe/H], since a star's convection zone is deeper for larger [Fe/H] \citep{mr02}. For each star in our samples we have T$_{\rm eff}$, $\log$ g, [Fe/H] and M$_{\rm v}$, all corrected for small systematic differences among the various studies.

Isochrone-based stellar ages can be derived from the T$_{\rm eff}$, [Fe/H] and M$_{\rm v}$ values, and this has been done in several studies of SWPs (e.g., \citet{tak207}). Uniform age estimates are not available for all the stars in our SWP and comparison star samples. However, our method of comparison does not require that we calculate ages. Instead, we calculate an index that is a measure of the proximity of two stars in T$_{\rm eff}$, $\log$ g, [Fe/H], M$_{\rm v}$-space. It is based on the following equation:

\begin{eqnarray*}
\Delta_1=30~\vert \log~{\rm T}_{\rm eff}^{\rm comp} - \log~{\rm T}_{\rm eff}^{\rm p} \vert +
\vert {\rm [Fe/H]}^{\rm comp} - {\rm [Fe/H]}^{\rm p} \vert \\
 + 0.5~\vert \log {\rm g}^{\rm comp} - \log {\rm g}^{\rm p} \vert + \vert {\rm M}_{\rm v}^{\rm comp} - 
 {\rm M}_{\rm v}^{\rm p} \vert
\end{eqnarray*}

Two stars with identical values of T$_{\rm eff}$, $\log$ g, [Fe/H] and M$_{\rm v}$ will have a $\Delta_1$ value of zero. Therefore, two stars with identical ages, masses and compositions will also have a $\Delta_1$ value of zero. Our motivation in choosing these particular coefficients is to give comparable contributions to $\Delta_1$ from each term for two stars that differ by about 2$\sigma$ in each parameter. We increased the contribution of the $\log$ T$_{\rm eff}$ term, however, to account for its greater importance in determining the Li abundance. And, we reduced the contribution of $\log$ g, given the relatively lesser importance of this quantity.

We calculated $\Delta_1$ for each SWP relative to each comparison star, resulting in 5439 values. For each SWP, we then selected the comparison star with the smallest value of $\Delta_1$. We plot the differences in Li abundances between each SWP and the closest matching comparison star in Figure 3a.

One weakness of this approach is that only a small subset of the data is used to calculate the differences in the Li abundances. These results are sensitive to outliers. We can improve upon our analysis as follows. In our second approach we calculated weighted average values of Li abundance differences for each SWP using all the comparison star data. The weights are given by $(\Delta_1)^{-2}$. We show the results of this analysis in Figure 3b.

The low Li abundances evident for the SWPs near 5800 K in Figure 3a are also evident in Figure 3b. The average Li abundance difference for SWPs with $5800 < $T$_{\rm eff} < 5950$ K is $-0.38 \pm 0.10$~(s.e.m.) dex. Figure 3b also shows a general excess of Li among SWPs with T$_{\rm eff} > 5950$ K, which have a mean Li abundance excess of $0.12 \pm 0.04$~(s.e.m.) dex.

To test the sensitivity of our results to the form of the $\Delta$ index, we repeated the above analysis with the following modified version:

\begin{eqnarray*}
\Delta_2=40~\vert \log~{\rm T}_{\rm eff}^{\rm comp} - \log~{\rm T}_{\rm eff}^{\rm p} \vert +
\vert {\rm [Fe/H]}^{\rm comp} - {\rm [Fe/H]}^{\rm p} \vert \\
 + 0.5~\vert \log {\rm g}^{\rm comp} - \log {\rm g}^{\rm p} \vert +1.5~\vert {\rm M}_{\rm v}^{\rm comp} - 
 {\rm M}_{\rm v}^{\rm p} \vert
\end{eqnarray*}

\begin{figure}
  \includegraphics[width=3.5in]{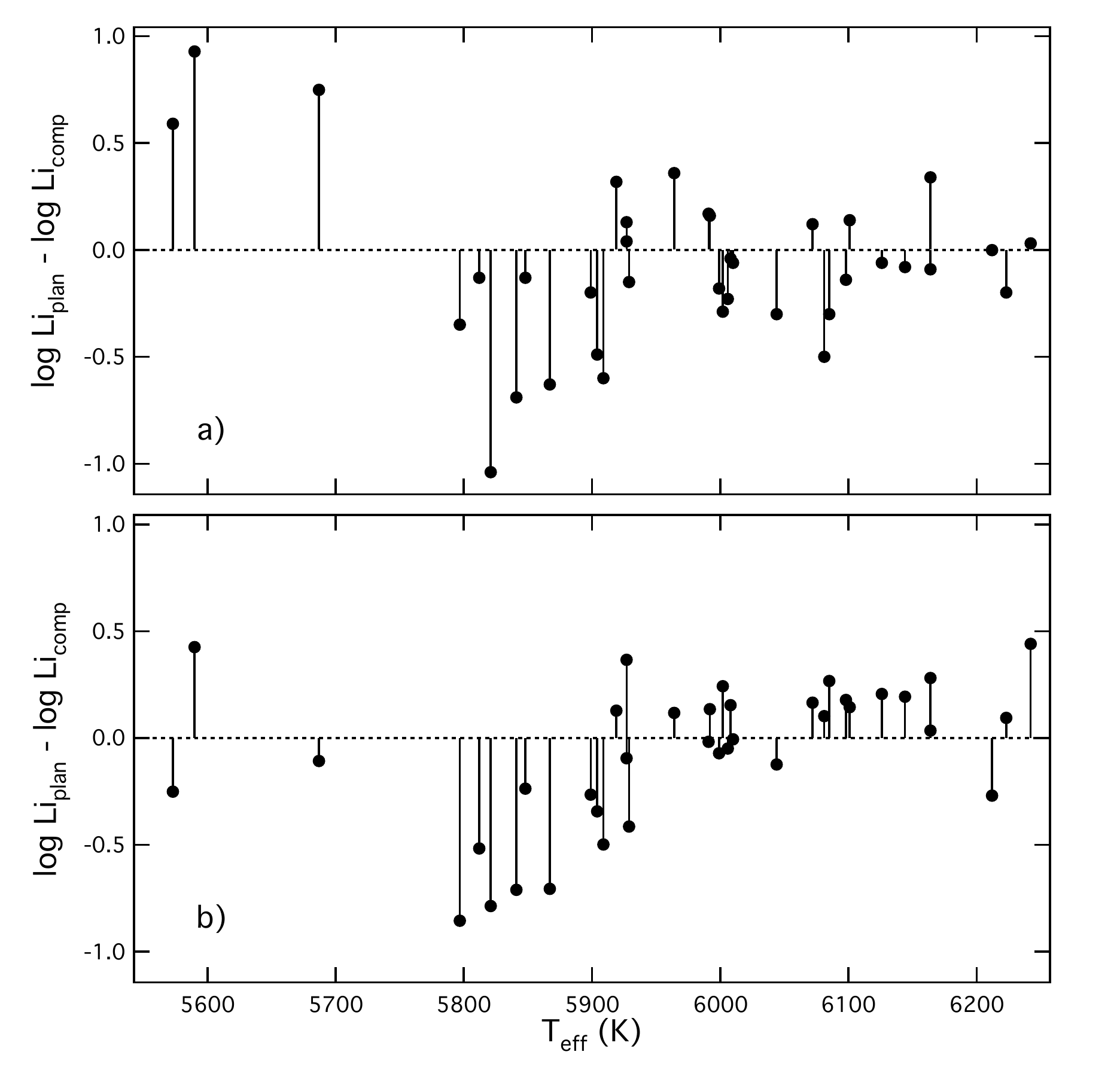}
 \caption{Li abundance differences (SWPs - comparison stars) versus T$_{\rm eff}$. Panel a shows the differences between SWPs and the most similar comparison stars. Panel b shows the weighted average differences. See text for additional details.}
\end{figure}

\begin{figure}
  \includegraphics[width=3.5in]{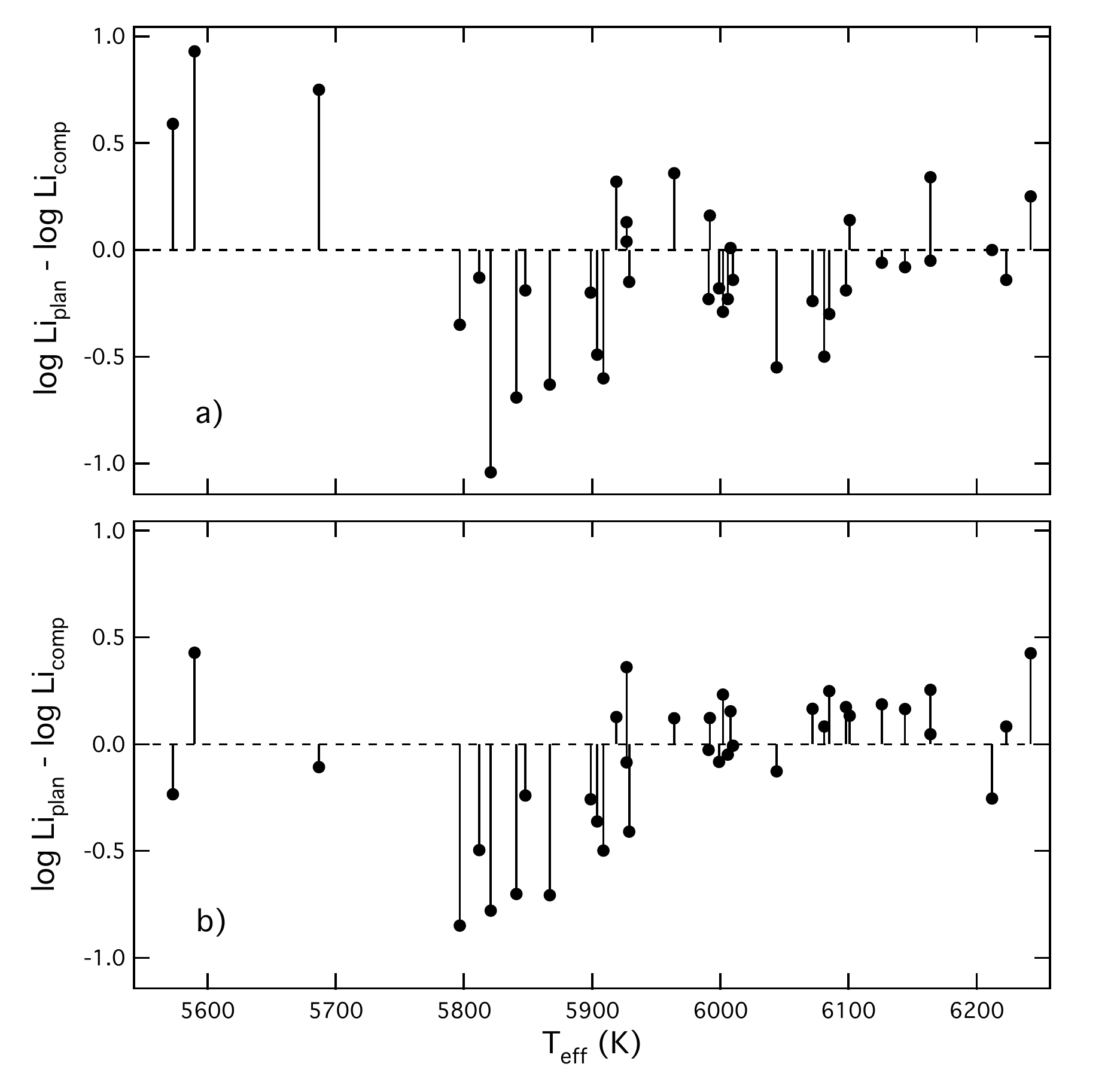}
 \caption{Same as Figure 3 but using the $\Delta_2$ index.}
\end{figure}

We show the resulting Li abundance differences calculated using the $\Delta_2$ index in Figure 4. There are only a few differences from Figure 3, but the pattern of low Li abundances among the cooler SWPs and the high Li abundances among the hotter SWPs evident in Figure 3b remain unchanged. The averages and dispersions are nearly identical to those calculated with the $\Delta_1$ index.

Rotation is another important parameter that correlates with Li abundance \citep{cut03}. Uniform rotational velocity (vsini) measurements are available from a single source for most of the stars we plotted in Figures 3 and 4. \citet{vf05} measured vsini for the 1040 stars in their SPOCS survey; they quote a typical uncertainty of 0.5 km s$^{\rm -1}$.\footnote{\citet{vf05} assigned a value of 0 km s$^{\rm -1}$ to 70 stars in their SPOCS database; 31 one of them pass our T$_{\rm eff}$ cuts. We assigned a value of 0.3 km s$^{\rm -1}$ to these stars prior to calculating the $\Delta$ indices.} This level of precision is adequate to resolve the range of vsini values measured for SWPs ($< 1$ to $\simeq 15$ km~s$^{\rm -1}$). Of the 37 SWPs plotted in Figures 3 and 4, 33 have vsini values listed in \citet{vf05}; in addition, 70 of the 147 stars in our comparison sample have vsini values listed in their study. We define a new $\Delta$-index that includes the vsini term:

\begin{eqnarray*}
\Delta_3=30~\vert \log~{\rm T}_{\rm eff}^{\rm comp} - \log~{\rm T}_{\rm eff}^{\rm p} \vert +
\vert {\rm [Fe/H]}^{\rm comp} - {\rm [Fe/H]}^{\rm p} \vert \\
 + 0.5~\vert \log {\rm g}^{\rm comp} - \log {\rm g}^{\rm p} \vert + \vert {\rm M}_{\rm v}^{\rm comp} - 
 {\rm M}_{\rm v}^{\rm p} \vert \\
 + \vert \log~{\rm vsini}^{\rm comp} - \log~{\rm vsini}^{\rm p} \vert
\end{eqnarray*}

We show the Li abundance differences calculated using the $\Delta_3$ index in Figure 5; the pattern apparent in the previous figures is still present. The average Li abundance difference for SWPs with $5800 < $~T$_{\rm eff} < 5950$ K is $-0.31 \pm 0.10$~(s.e.m.) dex; SWPs with T$_{\rm eff} > 5950$ K have an average Li abundance difference of $0.15 \pm 0.05$~(s.e.m.) dex.

\begin{figure}
  \includegraphics[width=3.5in]{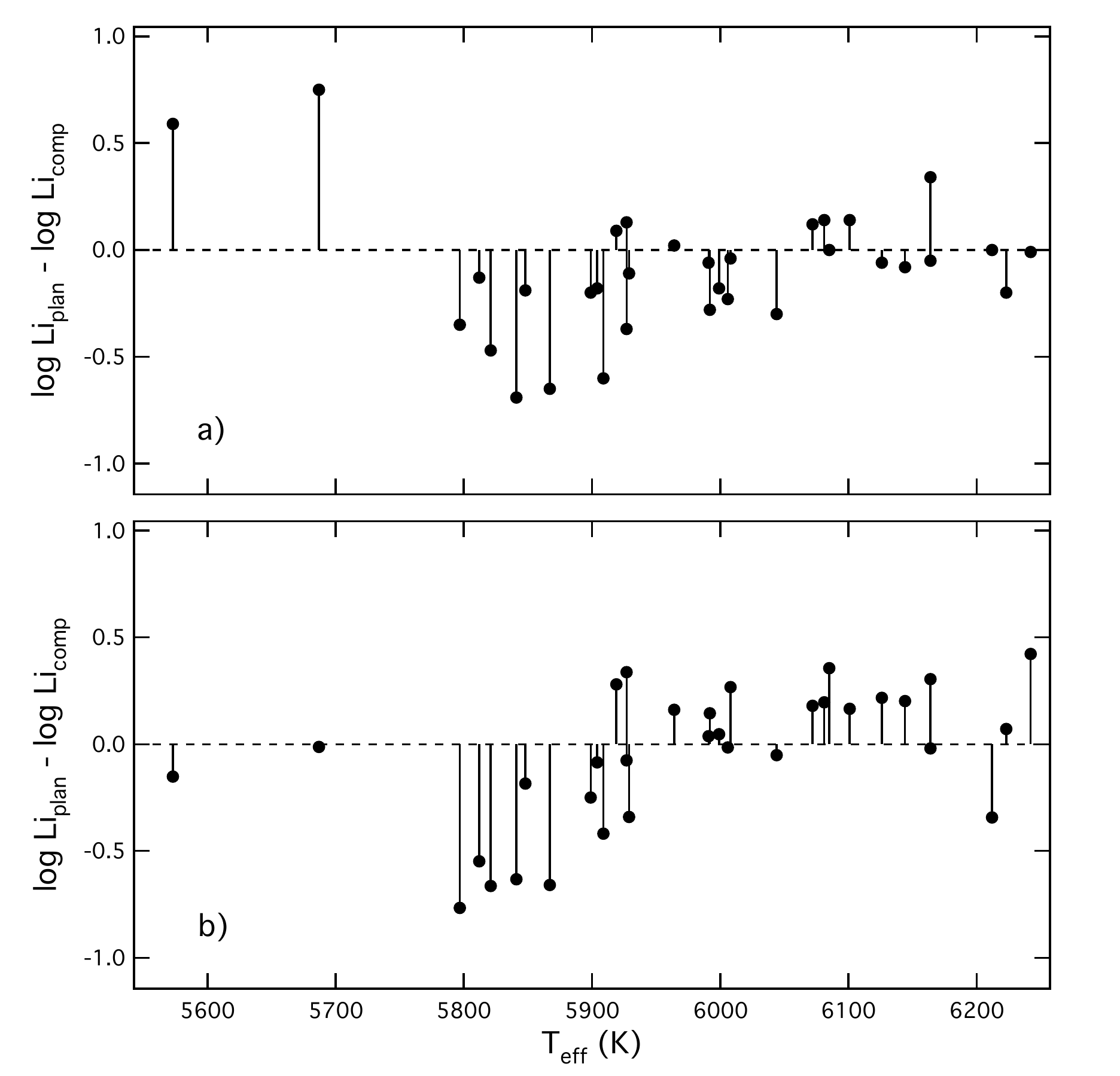}
 \caption{Same as Figure 3 but using the $\Delta_3$ index.}
\end{figure}

We repeated the calculations using an index with the vsini term three times the weight of that in the $\Delta_3$ index. The resulting Li abundance differences are very similar to those in Figure 5.

We summarize the properties of the SWP and comparison stars plotted in Figures 3 to 5 in Table 1.

\begin{table*}
\centering
\begin{minipage}{150mm}
\caption{Properties of SWPs and comparison stars plotted in Figures 3 to 5.}
\label{xmm}
\begin{tabular}{lccccccccc}
\hline
Figure & N & T$_{\rm eff}$ & T$_{\rm eff}$ & [Fe/H] & [Fe/H] & M$_{\rm v}$ & M$_{\rm v}$
& vsini & vsini\\
 & & (K) & (K) & (dex) & (dex) & (mags) & (mags) & (km~s$^{\rm -1}$) & (km~s$^{\rm -1}$)\\
~~Group & & range & mean & range & mean & range & mean & range & mean\\
\hline
3,4 & & & & & & & & &\\
~~SWP & 37 & 5573 - 6242 & 5972 & -0.69 - 0.37 & 0.10 & 3.32 - 5.36 & 4.16 & -- & --\\
~~comp. & 147 & 5550 - 6235 & 5890 & -0.87 - 0.39 & -0.01 & 2.72 - 5.53 & 4.56 & -- & --\\
5 & & & & & & & & &\\
~~SWP & 33 & 5573 - 6242 & 5978 & -0.69 - 0.37 & 0.10 & 3.32 - 4.82 & 4.09 & 1.3 - 9.6 & 3.6\\
~~comp. & 70 & 5550 - 6212 & 5898 & -0.27 - 0.31 & 0.04 & 3.25 - 5.39 & 4.52 & 0.3 - 9.8 & 3.4\\
\hline
\end{tabular}
\end{minipage}
\end{table*}

\section{Discussion}

We confirm recent claims that the Li abundances of SWPs with T$_{\rm eff}$ near 5800 K tend to be lower than those of stars without detected planets. We also find, for the first time, evidence that SWPs hotter than about 5900 K have excess Li. The magnitudes of the average the Li abundance differences for the cool and hot SWPs are larger than the typical measurement errors. We cannot reach any firm conclusions regarding SWPs cooler than 5800 K, given the paucity of such stars with detectable Li in our sample.

While our method of analysis does not rely on stellar evolutionary models, it does permit us to minimize the effects of differences in age and mass in our analysis of the Li abundances of the SWP and comparison star samples. In addition, the trends apparent in Figures 3 and 4 persisted even after correcting for differences in vsini. Inclusion of vsini as a parameter reduced the magnitude of the average Li abundance deficit among the cool SWPs only by 0.07 dex. 

Although the effect is small, vsini apparently does have an effect on the Li abundances of SWPs. To further explore the relationship between vsini and the presence of planets, we calculated vsini differences using the $\Delta_1$ index. We show the results for the same SWPs plotted in Figure 5 in Figure 6. Figure 6b shows that the vsini values are smaller than those of the comparison stars among the cooler stars; they are nearly the same for SWPs near 6000 K and much larger for the hottest ones.

\begin{figure}
  \includegraphics[width=3.5in]{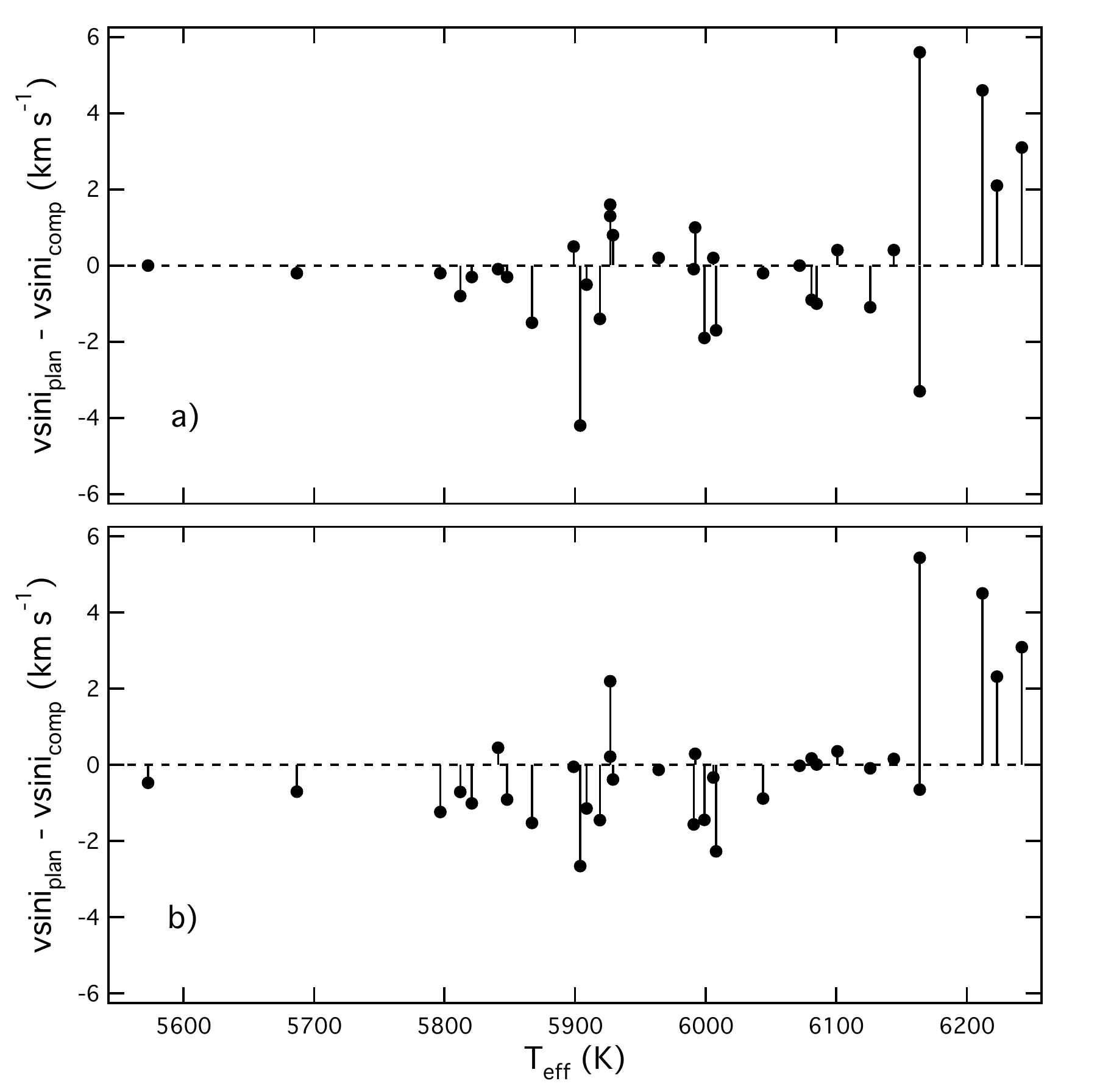}
 \caption{Plot of vsini differences versus T$_{\rm eff}$ for the same SWPs plotted in Figure 5.}
\end{figure}

We can improve our comparison of vsini values by employing the SWP and comparison star samples from \citet{vf05}. Applying the same T$_{\rm eff}$ and $\log$ g limits we used to prepare our samples above, we produced a samples of 82 SWPs and 596 comparison stars.\footnote{Since \citet{vf05} was published, several stars originally in their comparison stars list have been found to host planets. We have moved such stars to the SWP sample.} We show the resulting vsini differences calculated in the same way as in Figure 6 in Figure 7. These larger samples confirm our results from Figure 6. From the data in Figure 7b, we find that SWPs with T$_{\rm eff}$ between 5800 and 5950 K have an average vsini value of $-0.77 \pm 0.19$~(s.e.m.) km~s$^{\rm -1}$ relative to the comparison stars; the corresponding average difference for stars hotter than 5950 K is $+0.45 \pm 0.39$~(s.e.m.) km~s$^{\rm -1}$.

\begin{figure}
  \includegraphics[width=3.5in]{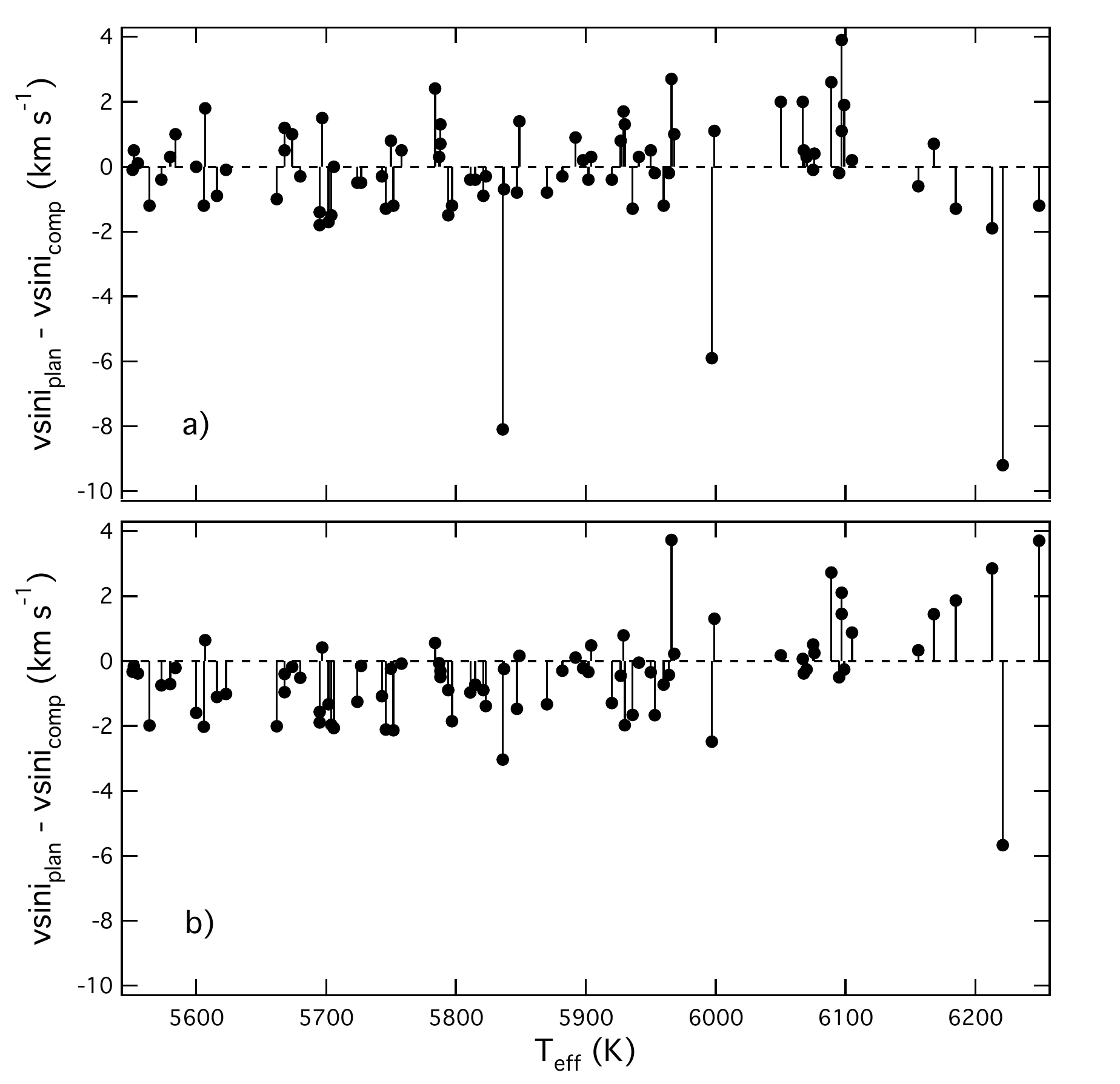}
 \caption{Plot of vsini differences versus T$_{\rm eff}$ for SWPs and comparison stars from \citet{vf05}.}
\end{figure}

The pattern of Li abundance differences between SWPs and comparison stars shows a remarkable correspondence with the pattern of differences in vsini. Both Li and vsini are smaller for cool SWPs and larger for hot SWPs. These results confirm the preliminary finding of \citet{tak07} that solar analogs with slower rotation have lower Li abundances. They reached this conclusion after having conducted detailed spectroscopic analyses of 118 solar analogs. They measured line widths, which allowed them to determine vsini+macroturbulence for each star. Unlike \citet{vf05}, however, they did not determine separate vsini and macroturbulence velocities. What's more, they only included 5 SWPs in their sample, limiting their conclusions about the possible relationship between SWPs, rotation and Li abundance.

Stellar chromospheric activity is known to correlate with rotation. If the pattern in vsini values we found among the SWPs is a real effect, then there should be similar patterns in measures of chromospheric activity. We can test this prediction using the $R^{'}_{\rm HK}$ chromospheric activity index, which has been measured for many Sun-like stars. The best source of $R^{'}_{\rm HK}$ values for our purposes is \citet{w04}. They tabulate $R^{'}_{\rm HK}$ values for the stars included in the \citet{vf05} study. In our first cut of the data, we cross referenced the stars in \citet{w04} with those in \citet{vf05} to produce new samples of SWPs and comparison stars. After applying our T$_{\rm eff}$ and $\log$ g cuts, we produced samples of 52 SWPs and 411 comparison stars. We show the resulting $R^{'}_{\rm HK}$ differences in Figure 8, calculated using the $\Delta_1$ index. As expected from the vsini results, SWPs tend to have smaller $R^{'}_{\rm HK}$ values. From the results shown in Figure 8b, we calculate that the average difference in $\log R^{'}_{\rm HK}$ for SWPs with T$_{\rm eff}$ between 5800 and 5950 K is $-0.10 \pm 0.02$~(s.e.m.) dex.

\begin{figure}
  \includegraphics[width=3.5in]{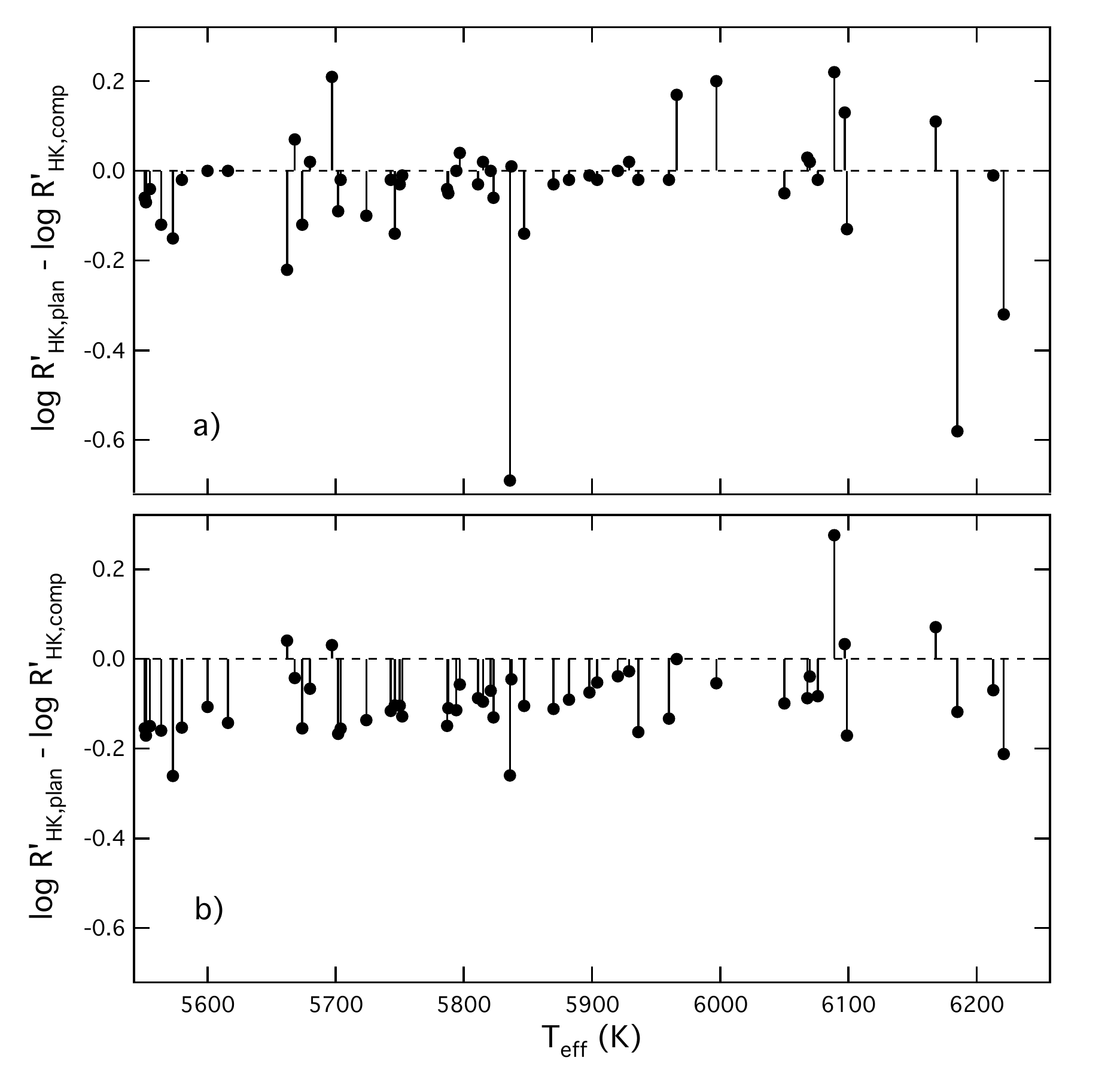}
 \caption{Plot of $\log R^{'}_{\rm HK}$ differences versus T$_{\rm eff}$ for SWPs and comparison stars from \citet{vf05} and \citet{w04}.}
\end{figure}

We should note that some scatter in the vsini and $R^{'}_{\rm HK}$ difference plots is expected. First, the equatorial velocity of a star is viewed from random angles. Second, like the Sun most FGK dwarfs display variable $R^{'}_{\rm HK}$.

Before we can attribute these results to physical causes, we need to consider the possible effects of biases. One important bias is the variation in the S/N ratio of the spectra, which we already discussed in Section 2. We dealt with it by excluding stars with Li abundances below the level where they could be uniformly determined.

Other biases are caused by differences in vsini and chromospheric activity, which affect the detectability of planets with the Doppler method (all the SWPs included in our comparisons above were discovered with the Doppler method). A large vsini value results in broad lines, and a large value of $R^{'}_{\rm HK}$ is accompanied by greater photospheric ``jitter," which is a source of noise in Doppler measurements \citep{w05}. The selection criteria of planet search groups includes a requirement that the spectra be sufficiently sharp-lined to detect the small Doppler shifts induced by an orbiting planet. Spectroscopic analyses of the type used to derive the basic stellar parameters (including Li abundance) also require sharp-lined spectra. Therefore, both the SWP and comparison star samples exclude stars with large vsini and $R^{'}_{\rm HK}$ values.

In order to test for possible SWP selection bias resulting from stellar activity, we produced a second set of SWP and comparison stars samples. This time we excluded stars with $\log R^{'}_{\rm HK} > -4.43$ dex; this is the largest value of $\log R^{'}_{\rm HK}$ measured for an SWP (HD 22049; \citet{j06} quote a value of $-4.43$ dex, while \citet{w04} quote a value of $-4.51$ dex). This second set contains the same number of SWPs (HD 22049 is too cool to have been included in our SWP samples) and 393 comparison stars. We plot these data in Figure 9. The average difference in $\log R^{'}_{\rm HK}$ for the SWP data in Figure 9b with T$_{\rm eff}$ between 5800 and 5950 K is $-0.07 \pm 0.01$~(s.e.m.) dex. Therefore, the activity bias can only account for a small part of the difference in $R^{'}_{\rm HK}$ between SWPs and comparison stars.

It is interesting to note that the star with the largest positive anomaly in Figures 8b and 9b, HD 179949, is the best candidate for star-planet magnetic interactions. \citet{s03} reported significant synchronous enhancement of Ca II H and K emission with the planet in orbit around the star.

\begin{figure}
  \includegraphics[width=3.5in]{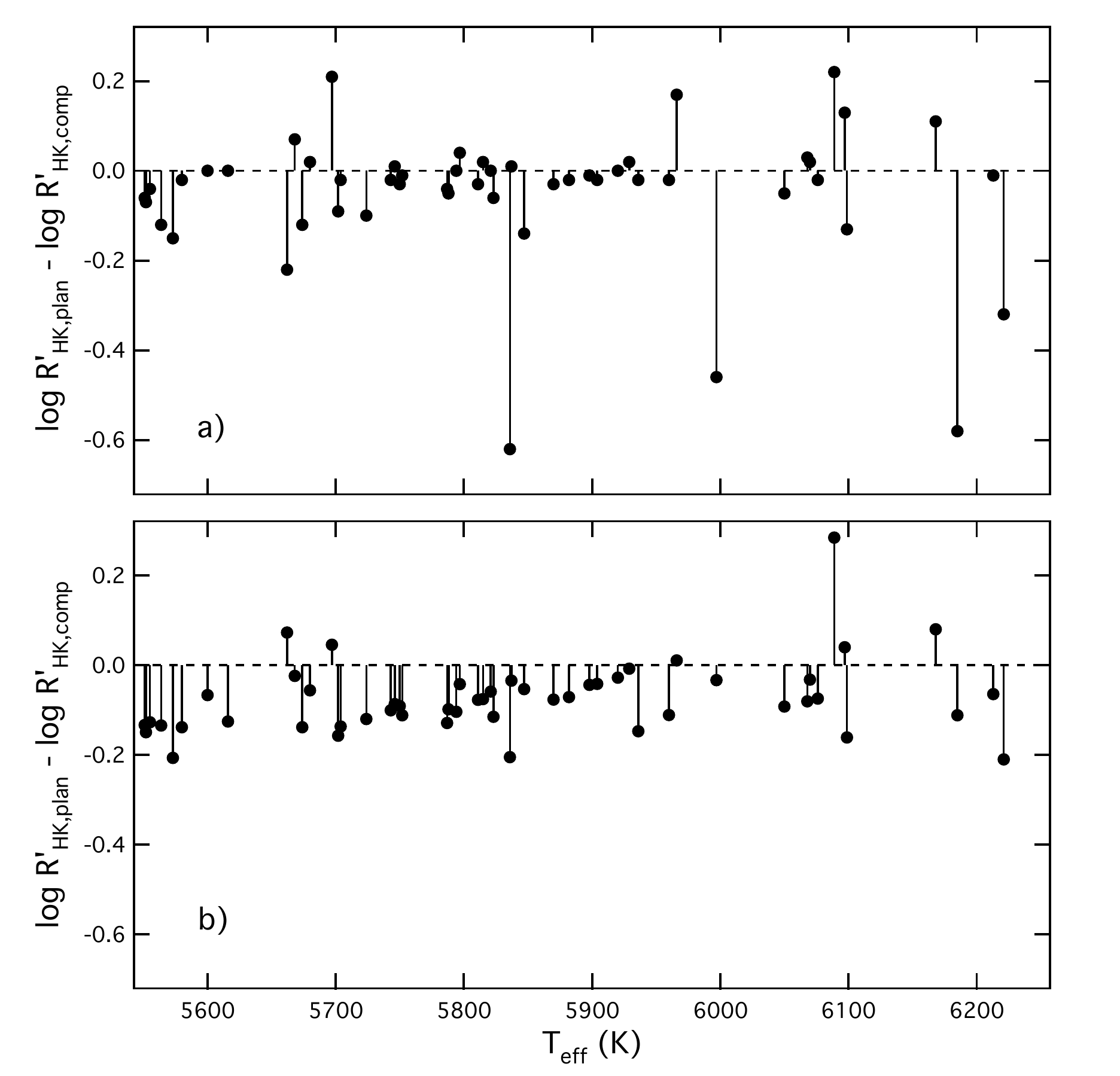}
 \caption{Same as Figure 8 but excluding stars with $\log R^{'}_{\rm HK} > -4.43$.}
\end{figure}

We also consider the vsini bias to have negligible effect on our results. First, the stars in the \citet{vf05} SPOCS database have median and maximum vsini values of 2.4 and 54.7 km~s$^{\rm -1}$, respectively, and only 44 stars have vsini values greater than 11 km~s$^{\rm -1}$. Only about 5 per cent of the comparison stars have vsini values greater than that of the SWP in the SPOCS sample with the largest vsini value (HD 120136 with vsini $= 15$ km~s$^{\rm -1}$). In addition, all the stars in the SPOCS database with vsini values above 16.5 km~s$^{\rm -1}$ are beyond the upper cutoff in T$_{\rm eff}$ (6250 K) that we employed in our sample preparations. 

Second, given that it would be more difficult to detect planets around stars with the highest vsini values, we would expect the SWPs sample to have a smaller average vsini value than a comparison sample that included stars with higher vsini values. However, we actually find that SWPs have positive vsini anomalies for T$_{\rm eff} > 6100$ K, the very range having the highest vsini values in the SPOCS database.

Third, vsini tends to decline with age. Therefore, if the SWP sample is significantly older than the comparison star sample, then the observed vsini values of the SWPs will be smaller than those in the comparison sample. We have greatly minimized this possible source of bias in our analysis above. By weighting the vsini and Li abundance differences according to the similarities of the stars, we are, in effect, comparing stars that are similar in age.

\citet{vf05} also excluded spectroscopic binaries and chromospherically very active stars. Again, we don't consider these biases to significantly affect our results for the reasons we gave above.

Are the Li abundance differences caused by differences in rotation or are they both caused by something else? The results shown in Figure 5b imply that vsini cannot fully explain the differences in Li abundance. The more likely explanation for the correlation between Li and rotation among SWPs is that both are the result of process related to planet formation.

This is not the first time Li abundance has been linked with rotation in stars. The low Li abundances observed over a narrow temperature range centered near spectral type F5 on the main sequence (called the Li dip) are accompanied by low vsini values; a similar phenomenon is observed among giants and subgiants at lower T$_{\rm eff}$ in the Hertzsprung gap. \citet{bv04} suggests that the simultaneous decrease of both Li and rotation velocity in these two classes of stars is caused by evolutionary changes leading to deep mixing, which destroys Li at the higher temperatures at the base of the envelope and redistributes angular momentum. In addition, Li is known to be preserved in short period synchronously rotating binaries, confirming stellar models that include rotationally induced mixing \citep{ryan95}.

Clearly, processes that affect mixing at the base of convection zone must be included in any theory offered to explain the correlation between Li abundance and rotation. \citet{is04} proposed two hypotheses to account for the low Li abundances they measured for SWPs with T$_{\rm eff}$ near 5800 K. First, they suggested that since the protoplanetary disk around a young star contains a large fraction of the system's angular momentum, it can cause rotational breaking of the star. The breaking, in turn, results in deeper mixing in the star and thus more efficient destruction of Li. Second, in a system with a migrating giant planet, tidal forces from the planet could create a shear instability at the interface between the convective and radiative zones, leading to more mixing and hence destruction of Li.

Our results for SWPs near 5800 K confirm the correlation between Li abundance and rotation based on the first hypothesis. Testing the second hypothesis would require examining the orbits of the planets as well. To be effective, such an analysis should employ a larger sample than ours.

Any theory must also account for the higher Li abundances and faster rotation we found for the SWPs near 6200 K. In this temperature regime the mass of the convection zone is much smaller. Such a star might have weaker coupling to the protoplanetary disk. In addition, a less massive convection zone would be more susceptible to alteration of the surface composition by accretion of refractory material \citep{gg98}.

\citet{lg03} and \citet{as05} argued that the anomalous chemical abundance pattern (including extremely high Li and Be abundances) of the early F dwarf J37 in NGC 6633 is best explained by accretion of chondritic material. This ``self-enrichment'' process could account for the Li excesses (though much smaller in magnitude) we found for the late F SWPs. In this context it is notable that \citet{sant04} found Be abundance excesses near 0.1 dex relative to comparison stars for SWPs between 6000 and 6200 K. This is comparable to the Li excesses we determined for the same stars. Unfortunately, they only observed two comparison stars in this temperature range, so their results should be considered as preliminary. Finally, accretion should also increase the $^{6}$Li abundance in a star's atmosphere, possibly to detectable levels \citep{r02,is03}.

The self-enrichment hypothesis could also account for the higher vsini values of the hotter SWPs. A migrating planet orbits its host star in the same direction as it rotates. As material falls onto the star (whether small planetary building blocks or planets), it adds angular momentum to the star's envelope. 

\section{Conclusions}

We prepared new SWP and comparison star samples by combining published spectroscopic data in a consistent way. Our analyses confirm those recent studies that have reported smaller Li abundances for SWPs near 5800 K. In addition, the data suggest that SWPs with T$_{\rm eff}$ near 6100 K have excess Li abundances. The transition occurs near T$_{\rm eff} = 5950$ K.

We also find that the patterns of vsini and $R^{'}_{\rm HK}$ anomalies with T$_{\rm eff}$ correlate with the Li anomalies; the vsini and $R^{'}_{\rm HK}$ values are smaller among the cooler SWPs.

The low Li abundance, vsini and $R^{'}_{\rm HK}$ values among the cool SWPs is consistent with the suggestion that a protoplanetary disk causes rotational breaking of its host star, leading to additional mixing in its envelope. The additional mixing, in turn, accelerates the destruction of Li in the star's envelope. On the other hand, the preliminary finding of high Li abundance and vsini values of the hot SWPs can be counted as evidence for the self-enrichment hypothesis. Additional measurements of Be and $^{\rm 6}$Li in SWPs and stars without planets near T$_{\rm eff} = 6100$ K can help us test the self-enrichment hypothesis for SWPs in this temperature range.

The number of SWPs with detected Li is still relatively small. The present situation can be improved by conducting detailed spectroscopic analyses on the more recently discovered SWPs. In addition, several published studies of SWPs are based on spectra of modest S/N ratio ($\sim 150$), preventing detection of Li in many of the cooler stars. It would be straightforward to obtain spectra of higher S/N ($\ge 300$) for these stars. It would also be helpful to observe additional stars without detected planets with comparable S/N ratio.

Our findings bring to five the number of stellar parameters that correlate with the presence of Doppler detected planets: metallicity, mass, Li abundance, vsini and $R^{'}_{\rm HK}$. There might also be additional secondary parameters, such as the $^{6}$Li/$^{7}$Li isotope ratio and Al/Fe and Si/Fe abundances ratios, but these require confirmation. Taken together, these five parameters should make it possible to select stars with a high probability of hosting planets.

\section*{Acknowledgments}

We thank the anonymous reviewer for helpful comments and suggestions.

\bsp

\label{lastpage}

\end{document}